\documentclass[epj]{mysvjour}
\usepackage{graphics}
\usepackage{graphicx}        
\usepackage{latexsym}        
\usepackage{amsfonts}        
\usepackage{amssymb}         
\usepackage{amsmath}         
\newcommand{\bc}{\begin{center}}
\newcommand{\ec}{\end{center}}
\newcommand{\be}{\begin{equation}}
\newcommand{\ee}{\end{equation}}
\newcommand{\bea}{\begin{eqnarray}}
\newcommand{\eea}{\end{eqnarray}}
\newcommand{\Dlr}{\overset{\leftrightarrow}{D}}
\newcommand{\Dl}{\overset{\leftarrow}{D}}
\newcommand{\Dr}{\overset{\rightarrow}{D}}

\def\msbar{\overline{\text{MS}}}

\def\ks{\kappa_{\text{sea}}}
\def\kv{\kappa_{\text{val}}}

\begin{document}
\title{Probing Nucleon Structure on the Lattice}
\author{M.~G\"ockeler\inst{1} \and
Ph.~H\"agler\inst{2} \and
R.~Horsley\inst{3} \and
Y.~Nakamura\inst{4} \and
D.~Pleiter\inst{4} \and
P.E.L.~Rakow\inst{5} \and
A.~Sch\"afer\inst{1} \and
G.~Schierholz\inst{4,6} \and
W.~Schroers\inst{4} \and
H.~St\"uben\inst{7} \and
J.M.~Zanotti\inst{3}\thanks{Presented by J.M.~Zanotti.
 at PAVI '06, Milos, Greece.
} \\
\emph{(QCDSF/UKQCD Collaboration)}
}                     
\institute{Institut f\"ur Theoretische Physik, Universit\"at
  Regensburg, 93040 Regensburg, Germany \and Institut f\"ur
  Theoretische Physik T39, Physik-Department der TU M\"unchen,
  85747~Garching, Germany\and School of Physics, University of
  Edinburgh, Edinburgh EH9 3JZ, UK \and John von Neumann-Institut
  f\"ur Computing NIC/DESY, 15738 Zeuthen, Germany \and Theoretical
  Physics Division, Dep.~of Math.~Sciences, University of Liverpool,
  Liverpool L69 3BX, UK \and Deutsches Elektronen-Synchrotron DESY,
  22603 Hamburg, Germany \and Konrad-Zuse-Zentrum f\"ur
  Informationstechnik Berlin, 14195 Berlin, Germany}
\date{Received: date / Revised version: date}
%
\abstract{
The QCDSF/UKQCD collaboration has an ongoing program to calculate
nucleon matrix elements with two flavours of dynamical ${\cal O}(a)$
improved Wilson fermions.
Here we present recent results on the electromagnetic form factors,
the quark momentum fraction $\langle x\rangle$ and the first three
moments of the nucleon's spin-averaged and spin-dependent generalised
parton distributions, including preliminary results with pion masses
as low as 320~MeV.
\PACS{
      {12.38.Gc}{Lattice QCD calculations}   \and
      {13.40.Gp}{Electromagnetic form factors}
     } 
} 
\maketitle
\section{Introduction}
\label{intro}
\vspace*{-3mm}

The ability of generalised parton distributions (GPDs) \cite{GPD} to
describe both exclusive and inclusive processes has led to an
enormous amount of interest in these functions both experimentally and
theoretically.
Not only do GPDs encompass the ordinary electromagnetic form factors
and parton distribution functions, but they also allow for the
computation of the total quark contribution to the nucleon spin
\cite{Ji} as well as revealing important information on the transverse
structure of the nucleon \cite{Diehl,Bu}.
A full mapping of the parameter space spanned by GPDs is an extremely
extensive task which needs support from non-perturbative techniques
like lattice simulations.

Substantial progress has already been made in computing the first
three moments of unpolarised, polarised \cite{QCDSF-1,MIT,MIT-2} and
tensor \cite{tensorGPDs} GPDs on the lattice.

In this paper we present recent results from the QCDSF /UKQCD
collaboration. In section~\ref{sec:emff} we investigate the $q^2$
dependence of the Dirac and Pauli electromagnetic form factors, while
section~\ref{sec:x} contains preliminary results for the average
fraction of the nucleon's momentum carried by the quarks, $\langle
x\rangle$.
Finally, in section~\ref{sec:gpds} we present results for the first
three moments of the GPDs $H$ and $\tilde{H}$.

\vspace*{-2mm}
\section{Electromagnetic form factors}
\label{sec:emff}
\vspace*{-2mm}

The study of the electromagnetic properties of hadrons provides
important insights into the non-perturbative structure of QCD.
The EM form factors reveal important information on the internal
structure of hadrons including their size, charge distribution and
magnetisation.
Phenomenological interest in these form factors has been revived by
recent Jefferson Lab polarisation experiments \cite{JLabFF} measuring
the ratio of the proton electric to magnetic form factors,
$\mu^{(p)}G_e^{(p)}(q^2)/G_m^{(p)}(q^2)$.
These experiments show that this ratio unexpectedly decreases almost
linearly with increasing $q^2$, indicating that the proton's electric
form factor falls off faster than the magnetic form factor.

A lattice calculation of the $q^2$ dependence of the proton's
electromagnetic form factors can not only allow for a comparison with
experiment, but also help in the understanding of the asymptotic
behaviour of these form factors.
Such a lattice calculation would also allow for the extraction of
other phenomenologically interesting quantities such as magnetic and
electric charge radii and magnetic moments.


\vspace*{-1mm}
\subsection{Lattice Techniques}
\vspace*{-1mm}

On the lattice, we determine the form factors $F_1(q^2)$ and
$F_2(q^2)$ by calculating the following matrix element of the
electromagnetic current
\begin{eqnarray}
\langle p',\,s'| j_{\mu}|p,\,s\rangle
 &=& \bar{u}(p',\,s') \bigg[ \gamma_\mu F_1(q^2) \nonumber \\
&&+\,
       i\sigma_{\mu\nu}\frac{q_\nu}{2m_N} F_2(q^2) \bigg] 
 u(p,\,s) \, ,
\label{eq:em-me}
\end{eqnarray}
where $u(p,\,s)$ is a Dirac spinor with momentum $p$ and spin
polarisation $s$, $q = p' - p$ is the momentum transfer with
$Q^2=-q^2$, $m_N$ is the nucleon mass and $j_\mu$ is the
electromagnetic current.

The form factors of the proton are obtained by using
\be
j_\mu^{(p)} = \frac{2}{3}\bar{u}\gamma_\mu u - \frac{1}{3}\bar{d}\gamma_\mu
d \ ,
\ee
while for iso-vector (i.e. proton $-$ neutron) form factors
\be
j_\mu^v = \bar{u}\gamma_\mu u - \bar{d}\gamma_\mu d \ .
\ee

It is common to rewrite the form factors $F_1$ and $F_2$ as 
\begin{eqnarray}
{\cal G}_e(q^2) &=& F_1(q^2) + \frac{q^2}{(2m_N)^2} F_2(q^2) , \\
{\cal G}_m(q^2) &=& F_1(q^2) + F_2(q^2) ,
\label{eq:Sachs}
\end{eqnarray}
which are known as the electric and magnetic Sachs form factors,
respectively. 

At zero momentum transfer, $F_1(0) = {\cal G}_e(0)$ gives the electric
charge (e.g. 1 for the proton), while
\be
{\cal G}_m^{(p)}(0) = \mu^{(p)} = 1 + \kappa^{(p)}\ ,
\ee
gives the magnetic moment, where $F_2^{(p)}(0) = \kappa^{(p)}$ is the
anomalous magnetic moment.

In order to extract the non-forward matrix elements from our lattice
simulations, we compute ratios of three- and two-point functions
\begin{eqnarray}
\label{eq:ratio}
\lefteqn{{\cal R}(t,\tau;\vec{p}\,',\vec{p};{\cal O})\, =
 \frac{C_\Gamma (t,\tau;\vec{p}\,', \vec{p},{\cal O})} 
        {C_2(t,\vec{p}\,')}} \\\nonumber
&&\times \left[
  \frac{C_2(\tau,\vec{p}\,') C_2(t,\vec{p}\,') C_2(t-\tau,\vec{p}\,)}
  {C_2(\tau,\vec{p}\,) C_2(t,\vec{p}\,) C_2(t-\tau,\vec{p}\,')}
\right]^{\frac{1}{2} }
\end{eqnarray}
which for large time separations, $0 \ll \tau \ll t \lesssim
\frac{1}{2} L_T $, where $L_T$ is the temporal extent of our lattice,
is proportional to the matrix element we are interested in, $\langle
p'| {\cal O}_q|p\rangle$.
The nucleon two- and three-point functions are given,
respectively, by
\bea
C_2(\tau,\vec{p}) &=& {\rm Tr}\left[\Gamma_{\rm unpol} 
\langle B(\tau,\vec{p}) \overline{B}(0,\vec{p})\rangle\right] \,,
\nonumber \\
C_\Gamma (t,\tau;\vec{p}\,', \vec{p},{\cal O}) &=& {\rm
  Tr}\left[\Gamma
\langle B(t,\vec{p}\,') {\cal O}(\tau)
\overline{B}(0,\vec{p})\rangle\right] \,. 
\label{eq:2-3ptfns}
\eea
Here $t$ and $\tau$ are the Euclidean times of the nucleon sink and
operator insertion, respectively, $\vec{p}\,'\ (\vec{p})$ is the nucleon
momentum at the sink (source), and ${\cal O}$ is the local vector current
\be
{\cal O}(\tau) = \psi(\tau)\gamma_\mu \psi(\tau)\ ,
\ee
which we renormalise non-perturbatively \cite{Bakeyev:2003ff}.
The trace in Eq.~(\ref{eq:2-3ptfns}) is over spinor indices and the
$\Gamma$ matrix determines the polarisation of the nucleon with
$\Gamma_\text{unpol}=\frac{1}{2}(1+\gamma_4)$.
%
We note here that in the calculation of nucleon matrix elements, we
neglect contributions coming from disconnected quark diagrams as these
are extremely computationally demanding.
Hence, in the following we mainly restrict ourselves to the
calculation of iso-vector matrix elements where the disconnected quark
contributions cancel.

Finally, we use the Sommer parameter, $r_0$, to set the scale with
$r_0=0.5$~fm.

\vspace*{-1mm}
\subsection{Results}
\vspace*{-1mm}

Of particular interest is the need to understand the behaviour of the
form factor $F_2(Q^2)$.
The question arises which is the best way to fit the form factor since
such a fitting function also allows an extrapolation of the form
factor to $Q^2=0$.
This is a necessary ingredient to find the anomalous magnetic moment
of the nucleon, $\kappa$.

Based on perturbative QCD, $F_1$ should scale asymptotically as
$1/Q^4$, while $F_2\sim 1/Q^6$ \cite{Brodsky:1974vy,Lepage:1980fj}.
It is difficult to obtain lattice data with high enough precision over
a large enough range of $Q^2$ values to distinguish between a dipole
or tripole behaviour.
It may, however, be instructive to consider the form factor ratio
$F_2(Q^2)/F_1(Q^2)$ since asymptotically this ratio should scale as
$1/Q^2$.
Spin polarisation experiments have instead found that the data is
compatible with
\be
\frac{F_2(Q^2)}{F_1(Q^2)} \sim \frac{1}{\sqrt{Q^2}}\ .
\ee

\begin{figure}[tb]
\bc
\vspace*{-20mm}
\includegraphics[angle=-90,width=9.92cm]{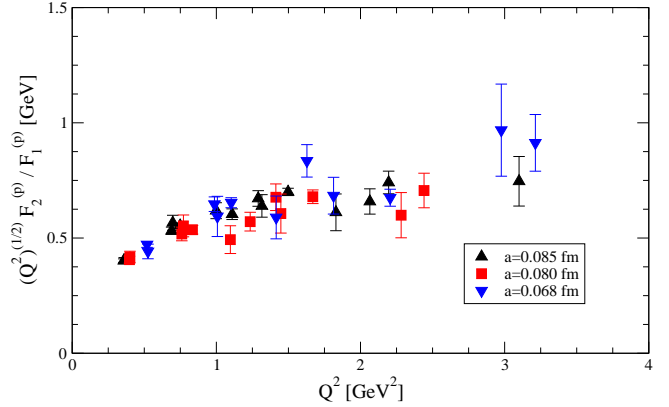}
\vspace*{-5mm}
\caption{$\sqrt{Q^2}\,F_2/F_1$ form factor ratio on three datasets with
  the same pion mass ($\approx 550$~MeV), but with different lattice
  spacings, $a=0.085,\,0.080,\,0.068$~fm.}
\label{fig:f2f1-const_mpi_sqrt}
\ec
\vspace*{-2mm}
\end{figure}

To investigate the asymptotic behaviour of the form factor ratio
$F_2(Q^2)/F_1(Q^2)$, we plot in Fig.~\ref{fig:f2f1-const_mpi_sqrt} the
results for $\sqrt{Q^2}F_2/F_1$ obtained at three working points with
approximately the same pion mass, but with different values of the
lattice spacing.
Here we observe the lattice data to be consistent with a constant for
$Q^2>1.5\,\text{GeV}^2$, similar to the experimental data.
Multiplying these results by an extra factor of $\sqrt{Q^2}$, as
suggested by perturbative QCD, would clearly destroy the plateau.
Quantitatively, though, the lattice data is higher than the
corresponding experimental ratios, cf \cite{Belitsky:2002kj}.
This shows that the lattice simulations are able to reproduce the
qualitative features of the experimental data, but for a quantitative
reproduction the pion mass is still unrealistically large.

In the following we fit $F_1$ and $F_2$ with a dipole ansatz

\be
F_i^{(v)}(q^2) = \frac{F_i(0)}{(1-q^2/M_i^2)^2}
\label{dipole}
\ee
where $F_1^{(v)}(0)=1$, $F_2^{(v)}(0)=\kappa^{(v)}$ and $M_i$ is the
fitted dipole mass for the form factor, $i$.

\begin{figure}
\resizebox{9cm}{!}{%
  \includegraphics{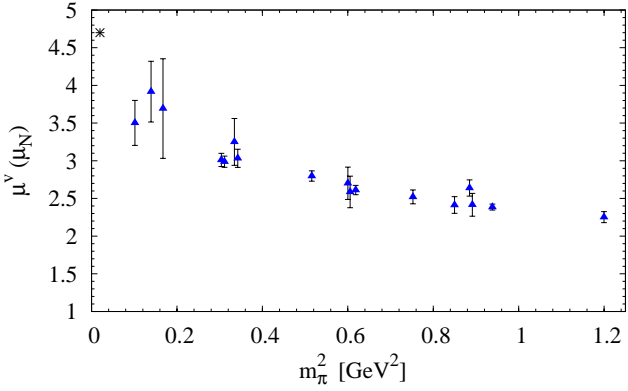}
}
\caption{Results for the isovector
  magnetic moment as a function of $m_\pi^2$. The experimental value
  is denoted by the star.}
\label{fig:MagMom}       
\vspace*{-1mm}
\end{figure}

We display our results for the isovector magnetic moment in
Fig.~\ref{fig:MagMom} as a function of $m_\pi^2$. 
Our results are in good agreement with recent quenched
\cite{Gockeler:2003ay,Boinepalli:2006xd,Alexandrou:2006ru} and $N_f=2$
\cite{Alexandrou:2006ru} results, which indicates that there appears
to be little effect due to quenching on the magnetic moments, as
predicted in \cite{Young:2004tb}.
The experimental value is indicated by a star at the physical pion
mass.
We clearly see that a linear extrapolation would miss the experimental
point.
This, however, is not completely unexpected as results from chiral
perturbation theory suggest that we should observe a dramatic increase
in the results at lighter pion masses
\cite{Gockeler:2003ay,Young:2004tb}.
The new points at lighter pion masses, $m_\pi^2<0.2$~GeV$^2$, are
beginning to show a hint of such curvature, although more work needs
to be done to reduce the error bars.

\vspace*{-2mm}
\section{Quark momentum fraction, $\langle x \rangle$}
\label{sec:x}
\vspace*{-2mm}

Forward matrix elements (no momentum transfer) provide moments of
quark distributions in some scheme, $\cal S$, at some scale, $M$:
\be
\langle N(\vec{p}) |
{\cal O}^{\{\mu_1\cdots\mu_n\}}_q |
N(\vec{p}) \rangle^{\cal S} = 
2 v_n^{(q){\cal S}}(g^{\cal S}(M))\;
p^{\mu_1}\cdots p^{\mu_n},
\ee
where 
\be
{\cal O}_q^{\{\mu_1\cdots\mu_n\}} = \overline{q} \, i^{n-1} \,
\gamma^{\{\mu_1}\, \Dlr^{\mu_2} \cdots
\Dlr^{\mu_n\}}\! q\ ,
\label{twist2}
\ee
$\Dlr = \frac{1}{2}(\Dr - \Dl)$ and $\{\cdots\}$ indicates
symmetrisation of indices and removal of traces.

Matrix elements with no momentum transfer are determined from a
simplified version of the ratio of three-point to
two-point correlation functions given in Eq.~(\ref{eq:ratio}).
See \cite{Gockeler:2004wp} for additional details.

We use non-perturbative renormalisation as outlined in Section~5.2.3
of \cite{Gockeler:2004wp} to convert our lattice results to the
$\msbar$ scheme at $\mu^2=4$~GeV$^2$.

In the language of the parton model, $v_n^q$ is often denoted by $\langle
x^{n-1} \rangle^q$
\be 
\langle x^{n-1} \rangle^q = \int^1_0 \, dx\, x^{n-1}\, [q(x) +
(-1)^n\bar{q}(x)] = v_n^q \ .
\ee

Of particular interest is the first ($n=2$) moment, $v_2^q = \langle
x\rangle^q $, which determines the fraction of the nucleon's momentum
carried by the quark, $q$.
This quantity is notorious on the lattice for producing values much
larger than phenominologically accepted results.
These discrepancies can possibly be explained by the fact that all
lattice calculations to date have been performed at quark masses that
are much larger than the physical masses \cite{Detmold:2001jb}.
Hence, it is a challenge for current lattice simulations to calculate
$\langle x\rangle$ at small enough quark masses in order to search for
the severe curvature predicted in Ref.~\cite{Detmold:2001jb}.

\begin{figure}
\resizebox{9cm}{!}{%
  \includegraphics{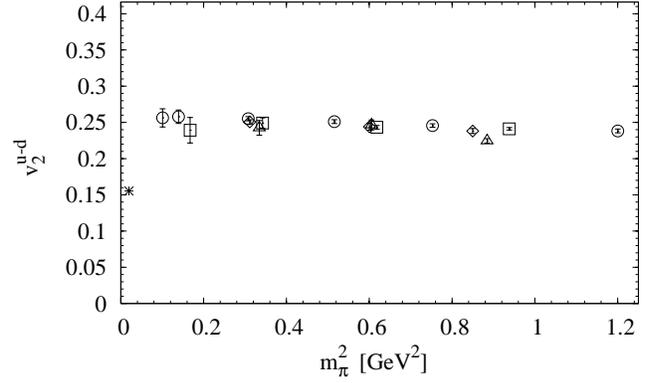}
}
\caption{Isovector $\langle x\rangle$ as a function of
  $m_\pi^2$ in the $\msbar$ scheme at $\mu^2=4$~GeV$^2$. These
  preliminary results are obtained at four different lattice spacings
  (in fm): 0.092 (triangles), 0.085 (diamonds), 0.080 (circles) and
  0.068 (squares). The star indicates the phenomenological
  result of the MRST analysis \cite{Martin:2001es} as given in
  \cite{Gockeler:2004wp}. This is in agreement with a recent higher
  order analysis \cite{Blumlein:2006be}.}
\label{fig:v2b}       
\vspace*{-1mm}
\end{figure}

Figure~\ref{fig:v2b} displays preliminary results for $\langle
x\rangle^{(u-d)}$ with pion masses as low as $\sim 320$~MeV. 
Before we can draw any conclusions on the behaviour at small quark
masses, we need to study scaling violations and finite size effects
more carefully.
Indeed, it has been suggested \cite{Detmold:2005pt,Detmold:2003rq}
that a volume of at least (4~fm)$^3$ is required to confirm the
predicted chiral curvature.

\vspace*{-2mm}
\section{Generalised parton distributions}
\label{sec:gpds}
\vspace*{-2mm}


\subsection{Matrix Elements And Moments of GPDs}
\vspace*{-1mm}

For a lattice calculation of GPDs, we work in Mellin-space
to relate matrix
elements of local operators to Mellin moments of the GPDs. 
The non-forward matrix elements of the twist-2 operator in
Eq.~(\ref{twist2}) specifies the $(n-1)^{th}$ moments of the
spin-averaged generalised parton distributions.
Replacing $\gamma$ with $\gamma_5\gamma$ leads to moments of the
spin-dependent GPDs.
In particular, for the unpolarised GPDs, we have
\vspace*{-3mm}
\be
\!\!\int_{-1}^1 dx\, x^{n-1}\, H^q(x,\xi, t) =
 H^q_{n}(\xi,t)\, ,
\!\int_{-1}^1 dx\, x^{n-1}\, E^q =
 E^q_{n}\, ,
\ee
where \cite{Ji}
\vspace*{-6mm}
\begin{eqnarray}
H^q_{n}(\xi,t) &=& \sum_{i=0}^{\lfloor\frac{n-1}{2}\rfloor}
 A^q_{n,2i}(t) (-2\xi)^{2i}
 + 
 C^q_{n}(t) (-2\xi)^n|_{n\text{ even}} \, ,\nonumber\\
E^q_{n}(\xi,t) &=& \sum_{i=0}^{\lfloor\frac{n-1}{2}\rfloor}
 B^q_{n,2i}(t) (-2\xi)^{2i}
 - 
 C^q_{n}(t) (-2\xi)^n|_{n\text{ even}} \,.\nonumber\\
\label{GPDmoments}
\end{eqnarray}
Here we denote the invariant of the momentum transfer by
$t=\Delta^2=(p'-p)^2$.
The generalised form factors $A^q_{n,2i}(t)$,
$B^q_{n,2i}(t)$ and $C^q_{n}(t)$ for the lowest three moments
are extracted from non-forward nucleon matrix elements of the
operators in Eq.~(\ref{twist2}) \cite{MIT}.

For the lowest moment, $A_{10}$ and $B_{10}$ are just the Dirac and
Pauli form factors $F_1$ and $F_2$, respectively
\begin{eqnarray}
\int_{-1}^1 dx\, H^q(x,\xi, t) &=& A^q_{10} (t) = F_1 (t)\, , \\
\int_{-1}^1 dx\, E^q(x,\xi, t) &=& B^q_{10} (t) = F_2 (t)\, ,
\end{eqnarray}
while $\widetilde{A}_{10}$ and $\widetilde{B}_{10}$ are the usual axial-vector
and pseudoscalar form factors, respectively
\begin{eqnarray}
\int_{-1}^1 dx\, \widetilde{H}^q(x,\xi, t) &=& \widetilde{A}^q_{10} (t) = g_A (t)\, , \\
\int_{-1}^1 dx\, \widetilde{E}^q(x,\xi, t) &=& \widetilde{B}^q_{10} (t) = g_P (t)\, .
\end{eqnarray}

We also observe that
in the forward limit ($t = \xi = 0$), the moments of $H_q$
reduce to the moments of the unpolarised parton distribution $A_{n0}(0)
= \langle x^{n-1}\rangle$.

\vspace*{-3mm}
\subsection{Results For Generalised Form Factors}
\label{sec:GFFresults}
\vspace*{-2mm}

\begin{figure}[t]
\bc
\includegraphics[width=8.7cm]{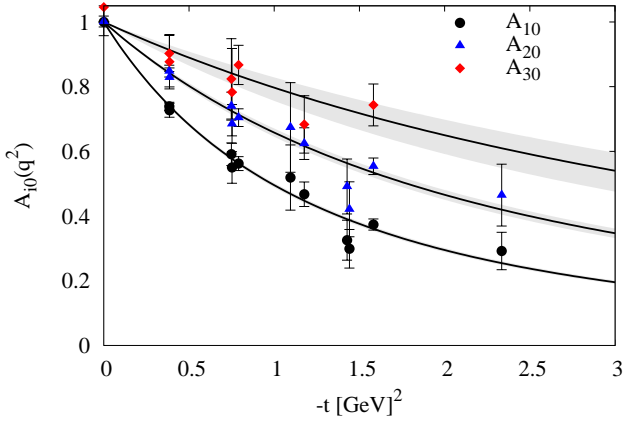}
\ec
\vspace*{-5mm}
\caption{Generalised form factors $A^{u-d}_{10},\ A^{u-d}_{20}$,
  $A^{u-d}_{30}$ together with a dipole fit. All form factors have
  been normalised to unity.}
\label{A123}
\vspace*{-5mm}
\end{figure}

\begin{figure}[t]
\bc
\includegraphics[width=8.7cm]{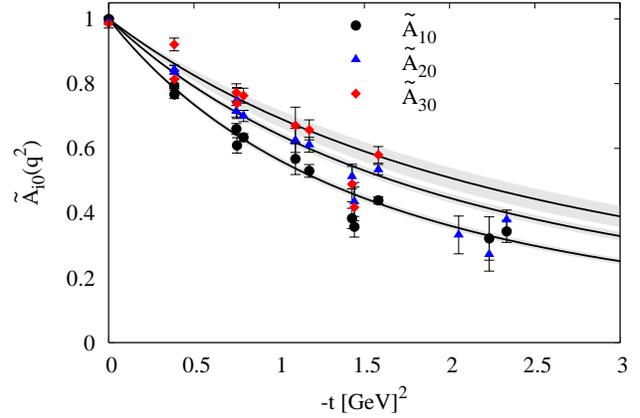}
\ec
\vspace*{-5mm}
\caption{Generalised form factors $\widetilde{A}^{u-d}_{10},\
  \widetilde{A}^{u-d}_{20}$, $\widetilde{A}^{u-d}_{30}$ together with
  a dipole fit. All form factors have been normalised to unity.}
\label{Apol123}
\vspace*{-5mm}
\end{figure}

Burkardt~\cite{Bu} has shown that the spin-independent and
spin-dependent generalised parton distributions $H(x,0,t)$ and
$\widetilde{H}(x,0,t)$ gain a probability interpretation when
Four-ier transformed to impact parameter space at
longitudinal momentum transfer $\xi=0$
\be
q(x,\vec{b}_\perp) =  \int \frac{d^2\Delta_\perp}{(2\pi)^2}\,
{\rm e}^{-i\vec{b}_\perp \cdot \vec{\Delta}_\perp}
H(x,0,-\Delta_\perp^2)\, ,
\label{IPGPD}
\ee
(and similar for the polarised $\Delta q(x,\vec{b}_\perp)$) where
$q(x,\vec{b}_\perp)$ is the probability density for a quark with
longitudinal momentum fraction $x$ and at transverse position (or
impact parameter) $\vec{b}_\perp$.

Burkardt \cite{Bu} also argued that $H(x,0,-\Delta_\perp^2)$ becomes
$\Delta_\perp^2$-independent as $x\rightarrow 1$ since, physically, we
expect the transverse size of the nucleon to decrease as $x$
increases,
i.e. $\lim_{x\rightarrow 1} q(x,\vec{b}_\perp) \propto
\delta^2(\vec{b}_\perp)$. As a result, we expect the slopes of the
moments of $H(x,0,-\Delta_\perp^2)$ in $\Delta_\perp^2$ to decrease as
we proceed to higher moments. 
This is also true for the polarised moments of
$\widetilde{H}(x,0,-\Delta_\perp^2)$, so from Eq.~(\ref{GPDmoments})
with $\xi=0$, we expect that the slopes of the generalised form
factors $A_{n0}(t)$ and $\widetilde{A}_{n0}(t)$ should
decrease with increasing $n$.

In Figs.~\ref{A123} and \ref{Apol123}, we show the $t$-dependence of
$A_{n0}(t)$ and $\widetilde{A}_{n0}(t)$, respectively,
$n=1,2,3$, for $\beta=5.40$, $\ks=\kv=0.13500$.
The form factors have been normalised to unity to make a comparison of
the slopes easier and we fit the form factors with a dipole form as in
Eq.~(\ref{dipole}).
We observe here that the form factors for the
unpolarised moments are well separated and that their
slopes do indeed decrease with increasing $n$ as predicted.
For the polarised moments, we observe a similar scenario, however here
the change in slope between the form factors is not as large.
The flattening of the GFFs $A_{n0}(t)$ has first been observed
in Ref.~\cite{MIT-2}, where at the same time practically no change in
slope was seen going from $\widetilde{A}_{20}(t)$ to
$\widetilde{A}_{30}(t)$.

Although fitting the form factors with a dipole is purely
phenomenological (see Ref.~\cite{Diehl:2004cx} for an alternative
ansatz), it does provide us with a useful means to measure the change
in slope of the form factors by monitoring the extracted dipole masses
as we proceed to higher moments.
We have calculated these generalised form factors on a subset of our
full complement of $(\beta,\, \kappa)$ combinations and have extracted
the corresponding dipole masses. 
Recall that $A_{10}$ is the Dirac form factor $F_1$, while
$\tilde{A}_{10}$ is the axial form factor $g_A$. Hence the dipole
fits can be compared with experiment.
A linear extrapolation produces a result larger than experiment for
both the polarised and unpolarised case, although the findings of
Ref.~\cite{Thomas} suggest that the chiral extrapolation of the dipole
masses of the electromagnetic form factors may be non-linear.

\begin{figure}[t]
\bc
\includegraphics[width=6.7cm]{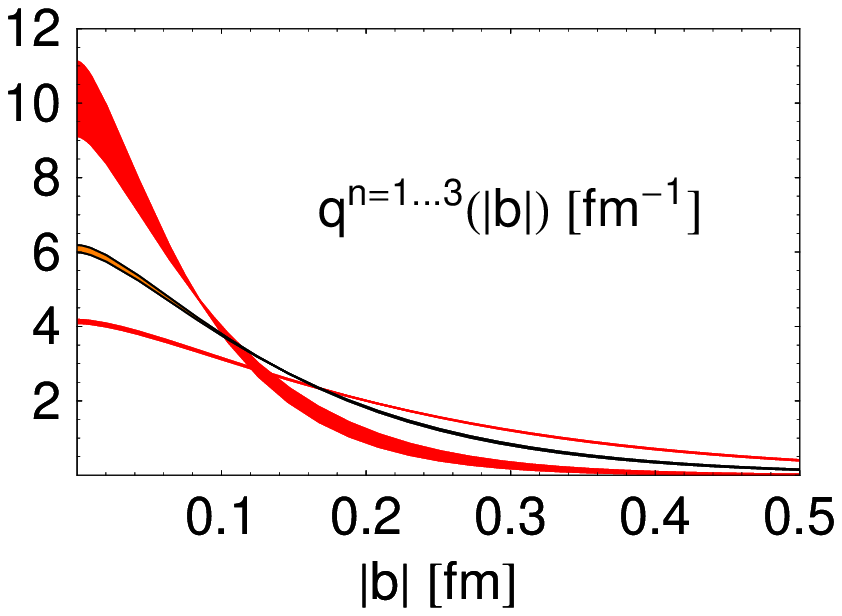}
\includegraphics[width=6.7cm]{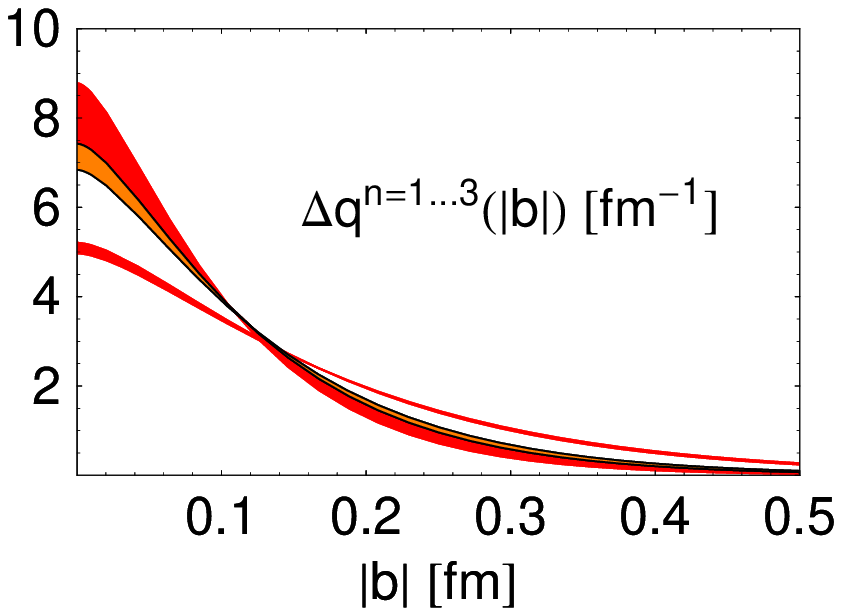}
\caption{The lowest three moments of the GPD $H(x,\xi=0,t)$
    (top) and $\widetilde{H}(x,\xi=0,t)$ (bottom) in impact parameter
    space as a function of impact parameter, $b$.}
\label{Density}
\ec
\vspace*{-5mm}
\end{figure}

In Fig.~\ref{Density} we show the lowest three moments of the GPD
$H(x,\xi=0,t)$ (top) and $\widetilde{H}(x,\xi=0,t)$ (bottom) in impact
parameter space.
The curves correspond to the Fourier-transformation of our dipole
ansatz Eq.~(\ref{dipole}), with the dipole masses extrapolated linearly
to the chiral limit, to $b_\perp$-space, and the shaded error
band is a result of the errors in the extrapolated dipole masses at
the physical pion mass.
The curves have been normalised so that they represent line densities
with $\int {\rm d}b\, q^n(b)=1$. 
The top figure of Fig.~\ref{Density} clearly shows how the $u-d$
quark distribution narrows as we proceed to higher moments $n$ and thereby
larger values of the average momentum fraction, while for the
polarised case in the bottom figure, the narrowing of the distribution
is not so severe.

\vspace*{-3mm}
\section*{Acknowledgements}
\vspace*{-2mm}

The numerical calculations have been performed on the Hitachi SR8000
at LRZ (Munich), the Cray T3E at EPCC (Edinburgh) \cite{Allton:2001sk}
the APE{\it 1000} and apeNEXT at NIC/DE-SY (Zeuthen), the BlueGeneL at
NIC/J\"ulich and the BlueGeneL at EPCC (Edinburgh). Some of the
configurations at the small pion mass have been generated on the Blue
GeneL at KEK by the Kanazawa group as part of the DIK research
programme.  This work was supported in part by the DFG, by the EU
Integrated Infrastructure Initiative Hadron Physics (I3HP) under
contract number RII3-CT-2004-506078. Ph.H. acknowledges support by the
DFG Emmy-Noether program.

\end{document}